\documentstyle[psfig,draft]{mn}

\def\gs{\mathrel{\raise0.35ex\hbox{$\scriptstyle >$}\kern-0.6em
\lower0.40ex\hbox{{$\scriptstyle \sim$}}}}
\def\ls{\mathrel{\raise0.35ex\hbox{$\scriptstyle <$}\kern-0.6em
\lower0.40ex\hbox{{$\scriptstyle \sim$}}}}

\title[UKIRT Archive Catalogue of AO Fields]
      {A Catalogue of Potential Adaptive Optics Survey Fields from the UKIRT Archive}

\author[Christopher \& Smail]
       {N.\,M.\ Christopher$^{1}$  \& 
	Ian Smail$^{2}$ 
        \vspace*{1mm}\\
        $^1$ Department of Physics, Durham University, South Road,
        Durham, DH1 3LE, UK\\        
	$^2$ Institute for Computational Cosmology, 
	Durham University, South Road,
        Durham, DH1 3LE, UK\\
}

\date{Accepted ... ; Received ... ; in original 2005 August 18}

\pagerange{000--000}

\begin{document}

\maketitle

\begin{abstract}
We present a multicolour catalogue of faint galaxies situated close to
bright stars, $V\ls 15$, with the aim of identifying high-redshift
galaxies suitable for study with adaptive optics-equipped near-infrared
imagers and spectrographs. The catalogue is constructed from archival
calibration observations of UKIRT Faint Standard stars with the UFTI
camera on UKIRT.  We have analysed the deepest 16 fields from the
archive to provide a catalogue of galaxies brighter than $K\sim 20.3$
lying within 25$''$ of the guide stars.  We identify 111 objects in a
total survey area of 8.7 sq.\ arcmin, of these 87 are classified as
galaxies based on their light profiles in our $\sim 0.5''$ median
seeing $K$-band images.  Of these, 12 galaxies have $(J-K)\geq 2.0$
consistent with them lying at high-redshifts, $z\gs 2$.  These 12 very
red galaxies have $K$-band magnitudes of $K=18.1$--20.1 and separations
from the guide stars of 4--20$''$ and hence are very well-suited to
adaptive optics studies to investigate their morphologies and spectral
properties on sub-kpc scales. We provide coordinates and $JHK$
photometry for all catalogued objects.
\end{abstract}


%
%
\begin{figure*}
\centerline{\psfig{file=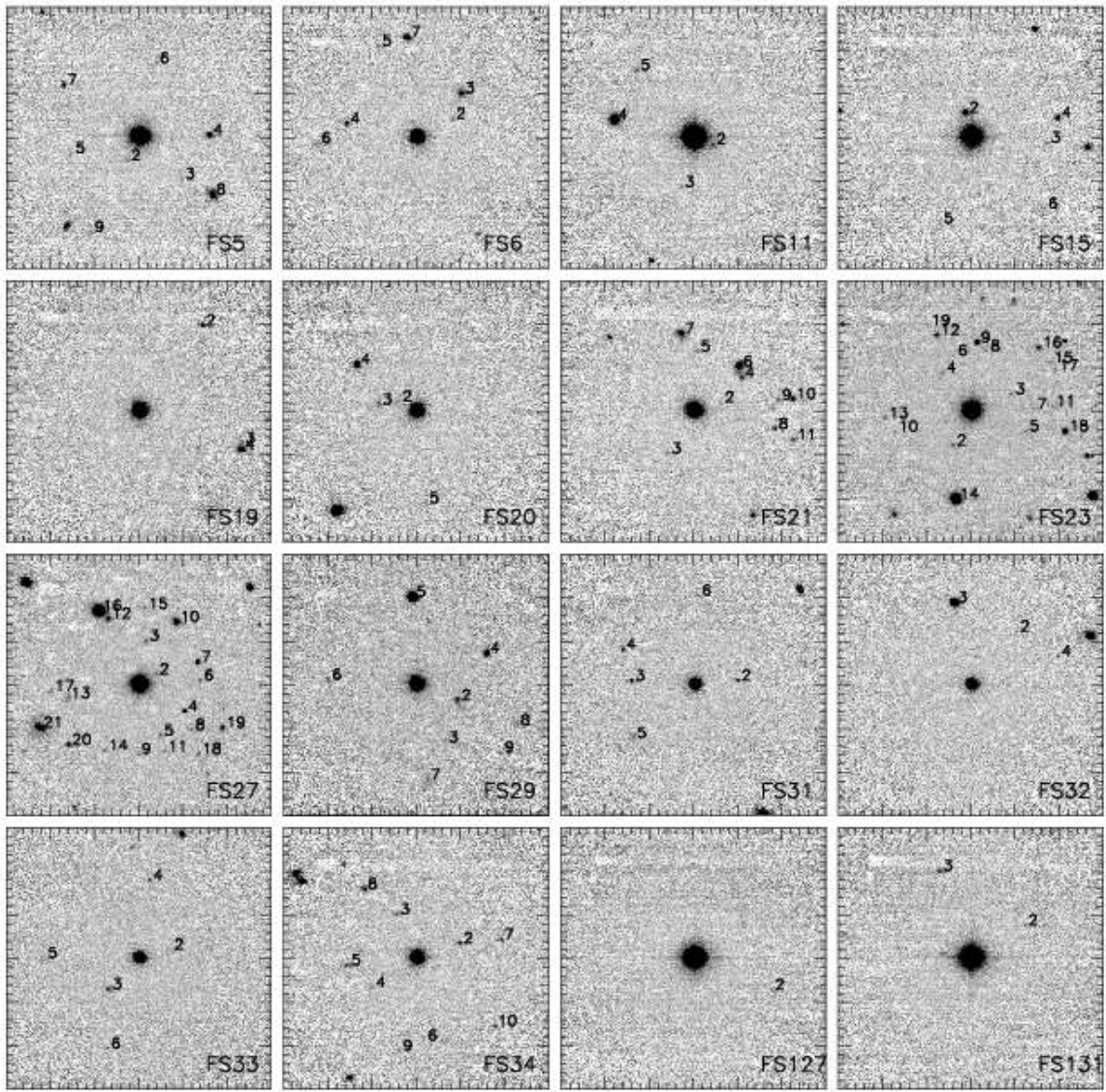,angle=0,width=6.0in}}
 \caption{The central 60$''\times$60$''$ region of the $K$-band images
of the 16 fields in our sample.  We identify the other objects in the
fields using the numbering scheme from Table~2. Each panel is centred on the faint
standard in the field and has North top and East to the left.  The
major tick marks on the axes are every 10$''$.}
\end{figure*}

\section{Introduction}

The high-resolution optical and near-infrared imaging from {\it Hubble
Space Telescope} ({\it HST}) has provided unique insights into the
morphological evolution of galaxies in the restframe optical over the
$\sim 8$\,Gyrs to $z\sim 1$ (e.g.\ Glazebrook et al.\ 1995; Giavalisco
et al.\ 2004; Thompson et al.\ 2001).  However, the warm thermal
environment of {\it HST} precludes sensitive imaging at wavelengths
beyond $\sim 1.5\mu$m (the middle of the $H$-band).  This means that
{\it HST} cannot track the restframe optical morphologies of galaxies
much past $z\sim 1$--1.5 and has to rely on imaging their restframe UV
emission -- a spectral region much more sensitive to recent star
formation and the influences of dust.

Fortunately, adaptive optics (AO) systems operating in the
near-infrared waveband on large ground-based telescopes can be used to
compliment the optical view provided by {\it HST} with comparable (or
better) spatial resolution: $\ls 0.1$--0.2 arcsec on faint galaxies
(Larkin et al.\ 2001; Glassman et al.\ 2002; Cresci et al.\ 2005).
These ground-based AO systems also provide the exciting opportunity to
obtain very high angular resolution spectroscopy of these galaxies,
necessary to study their kinematics on kpc scales (e.g.\ Eisenhauer et
al., 2003).  The drawback with such studies is the need for a bright
nearby star to provide a suitable reference source for wavefront
correction to remove the blurring caused by the atmosphere.  For
typical AO systems the guide star must be brighter than $V\sim 14$ and
situated within $\sim 20$ arcsec for it to provide any measurable
improvement to the achieved image quality (Le Louarn et al.\ 1998; Doel
et al.\ 2000; Benn 2001).  However, some correction can be achieved
with fainter guide stars or larger separations if these are
supplemented by information about the atmospheric phase transmission
from an additional artificial laser guide star.
 
The restriction on AO studies of faint galaxies is thus the need to
have a suitable bright guide star within the field.  This requirement
severely limits the number of suitable targets and has resulted in
adaptive optics studies so far having had relatively limited impact on
the study of faint galaxies, although the recent commissioning of laser
guide star capabilities on several 8--10-m class telescopes (e.g.\
Minowa et al.\ 2005) may change this.  To provide target fields
suitable for natural guide star studies, two groups have published
catalogues of galaxies identified close to bright stars: Larkin \&
Glassman (1999) and Baker et al.\ (2003).

Larkin \& Glassman provided $K$-band (and some $J$-/$H$-band
photometry) for around 40 galaxies to a 5-$\sigma$ limit of $K\sim 20$
in 5 patches close to $R=8.5$--10.3 bright stars in the Northern
hemisphere.  Due to the need to place the stars in one corner of the
array (to limit the effects of charge bleeding), all but $\sim 15$ of
these galaxies are beyond 25-arcsec from the correction star, beyond
which the corrections provided by typical AO systems using a single
star become negligible.  In contrast, Baker et al.\ (2003) imaged
regions around 42 $R=9.0$--12.4 guide stars in the Southern Hemisphere
in just the $K$-band to a 5-$\sigma$ limit of $K\sim 19.5$.  They find
around 350 galaxies within 25 arcsec of the guide stars.

Both of these previous studies have used targeted observations to
construct catalogues of sources around bright stars. Larkin \& Glassman
(1999) used 4 nights on the Keck telescope and the NTT took 6 nights to
complete the Baker et al.\ (2003) study. However, there are
opportunities of obtaining comparably deep data for free from
calibration fields on ground- (and indeed space-based) telescopes, by
exploiting archives now available for most major facilities.

Our study focuses on the archive for one telescope which is
particularly well-suited to this programme: the UK Infrared Telescope
(UKIRT).\footnote{The United Kingdom Infrared Telescope (UKIRT) is
operated by the Joint Astronomy Centre on behalf of the UK Particle
Physics and Astronomy Research Council.}  UKIRT operates exclusively in
the near- and mid-infrared -- the preferred spectral region for
ground-based AO studies of high-redshift galaxies.  The telescope also
has a fully-integrated data reduction pipeline associated with the
primary science instruments on the telescope and hence the reduction of
data from the archive can exploit the same pipeline to efficiently
process the large quantities of data necessary for the project.  The
bulk of the photometric calibration for the UKIRT imaging cameras also
relies on a small number of standard stars -- the UKIRT Faint Standards
(FS) from Hawarden et al.\ (2001).  The UKIRT FS list is ideal as the
stars in the primary list are predominantly equatorial and typically
bright enough (but not too bright) to use as AO guide stars.  The
majority of the recent imaging data are taken with a single instrument,
UFTI, guaranteeing it is homogeneous. There are approximately 140hrs of
observations of faint standard stars in the $JHK$ filters with UFTI in
the UKIRT archive.  The majority of these observations are of the
primary standard list from Hawarden et al.\ (2001).  This means we can
expect the cumulative integration on many of the calibration fields to
reach the depths necessary for studying the morphologies of $z\gs 1$
galaxies, i.e.\ $K\sim 20$.

We have therefore searched the UKIRT Archive for suitable FS fields,
reduced, calibrated and catalogued these to provide a list of faint
galaxies suitable for studies with current adaptive optics imagers and
spectrographs. This paper is structured as follows: in \S2 we detail
the field selection and the reduction of the data. In \S3 we describe
the analysis of these fields and discuss the properties of the sources
we detect. Finally, in \S4 we give our conclusions.

%
%
\begin{table*}
\begin{center}
\caption{Log of the observations of Faint Standards in the UKIRT
Archive.  We list the ID, $K$- and $V$-band magnitudes for the star,
its nominal position, total exposure times for the final stacked $JHK$
images, the FWHM of the $K$-band frame, the 5-$\sigma$ limiting
$K$-band magnitude (measured in a 2$''$-diameter aperture) and the
total number of stars and galaxies (excluding the FS) in our catalogues within
25$''$ of the FS and brighter than this limit.
}
\begin{tabular}{lccccrrrccrr}
\noalign{\medskip}
\hline
Star & $K$ & $V$ & R.A.\ & Dec.\ &  \multispan3{ ~T$_{\rm Exp}$ } & FWHM$_K$ &
$K_{\rm lim}$ & \multispan2{ ~~N } \\ 
    & & & \multispan2{ ~(J2000) } & $J$~ & $H$~ & $K$~ & ($''$) & 
& Star & Gal. \\
\noalign{\smallskip}
\hline
FS\,5 & 12.34 & 12.41 & 01 54 34.65 &$-$06 46 00.4 & 2450 &  850 &
2690 & 0.57 & 20.1 & 1 & 7 \cr 
FS\,6 & 13.38 & 12.80 & 02 30 16.64 &$+$05 15 51.1 & 1275 &  750 &
3525 & 0.52 & 20.4 & 0 & 6 \cr 
FS\,11 & 11.25 & 11.72 &04 52 58.92 &$-$00 14 41.6 & 1825 &  785 &
1775 & 0.72 & 19.7 & 2 & 2 \cr 
FS\,15 & 12.35 & 14.05 &08 51 05.81 &$+$11 43 46.9 & 4625 & 2130 &
3655 & 0.67 & 20.4 & 0 & 5 \cr 
FS\,19 & 13.78 & 13.01 &10 33 42.75 &$-$11 41 38.3 & 2375 & 1725 &
3525 & 0.78 & 20.3 & 1 & 2 \cr 
FS\,20 & 13.50 & 13.06 &11 07 59.93 &$-$05 09 26.1 & 2200 & 1300 &
2350 & 0.57 & 20.3 & 1 & 3 \cr 
FS\,21 & 13.15 & 12.50 &11 37 05.15 &$+$29 47 58.4 & 5300 & 2800 &
5275 & 0.55 & 20.7 & 0 & 10 \cr 
FS\,23 & 12.38 & 14.75 &13 41 43.57 &$+$28 29 49.5 & 5750 & 1905 &
4795 & 0.59 & 20.4 & 9 & 9 \cr
FS\,27 & 13.13 & 14.70 &16 40 41.56 &$+$36 21 12.4 & 7050 & 4025 &
9250 & 0.66 & 20.5 & 6 & 14 \cr 
FS\,29 & 13.31 & 12.74 &21 52 25.36 &$+$02 23 20.7 & 4950 & 4800 &
6425 & 0.56 & 20.5 & 1 & 7 \cr 
FS\,31 & 14.04 & 13.09 &23 12 21.60 &$+$10 47 04.1 & 2100 &  900 &
3650 & 0.54 & 20.4 & 1 & 4 \cr 
FS\,32 & 13.68 & 12.96 &23 16 12.37 &$-$01 50 34.6 & 1850 &  750 &
2650 & 0.52 & 20.3 & 2 & 1 \cr 
FS\,33 & 14.25 & 13.42 &12 57 02.30 &$+$22 01 52.8 & 2300 & 1750 &
3150 & 0.64 & 20.5 & 0 & 5 \cr 
FS\,34 & 13.00 & 12.34 &20 42 34.73 &$-$20 04 34.8 & 2050 &  775 &
1575 & 0.55 & 20.0 & 0 & 9 \cr 
FS\,127 & 14.20 & 13.86 &10 06 29.03 &$+$41 01 26.6 & 2315 & 775 &
2970 & 0.64 & 20.4 & 0 & 1 \cr 
FS\,131 & 11.75 & 12.56 &12 14 25.40 &$+$35 35 55.6 & 1975 & 855 &
2220 & 0.64 & 19.8 & 0 & 2 \cr 
\hline
\end{tabular}

\end{center}
\end{table*}

\section{Observations \&  Reduction}

The first step in our study was to query the UKIRT Archive\footnote{The
UKIRT Archive is provided through the UK Astronomy Data Centre,
http://archive.ast.cam.ac.uk/ukirt\_arch} for all the observations in
the $JHK$-bands with the UFTI camera (Roche et al.\ 2002) of stars in
the extended UKIRT Faint Standard List (Hawarden et al.\ 2001).  We
then select all those with potentially more than 2.5-ks integration in
the $K$-band and available in archive as of 2005 August 1st.  We
removed any FS with $V\gs 15$ (the deepest likely limit for a natural
guide star with present AO systems) and those with reddening in the
field of $A_K\geq 0.02$ ($A_V>0.2$) based on Schlegel et al.\ (1998) as
estimated from NED.\footnote{The NASA/IPAC Extragalactic Database (NED)
is operated by the Jet Propulsion Laboratory, California Institute of
Technology, under contract with the National Aeronautics and Space
Administration.}

~From the initial 83 faint standards (28 from the primary list, those
with IDs $<40$, and 55 from the extended list, those with IDs $>100$)
this leaves us with 17 fields.  We discard one field, FS\,16, due to
the combination of shallow depth and poorer than average image quality
-- leaving 16. We list the IDs, coordinates and total exposure times in
the $J$-, $H$- and $K$-bands for these fields in Table~1. 

Observations of standards taken with UFTI typically jitter the star
around the lower-right-hand quadrant of the HAWAII-1 detector and only
read-out this $512\times512$ pixel region.  At the nominal pixel scale
of UFTI, 0.0908\,arcsec\,pixel$^{-1}$, this field corresponds to $\sim
46''$ -- well matched to the isoplanatic patch in the near-infrared.
The standard calibration sequence consists of a dark exposure, followed
by 5 exposures of the standard in a diagonal cross pattern.

Data are reduced using the {\sc oracdr} pipeline for UFTI data
(Economou et al.\ 2004).  Due to the development of the pipeline and
changes in the file header parameters, slightly different incantations
are needed for data taken on different dates.  Prior to 2000 August 1
we used the {\sc oracdr\_ufti\_old} setup from the {\sc starlink}
v.2004.2 release, once the {\sc .fit} file extensions are renamed to
{\sc .fits} this automatically processes the data, subtracting suitable
dark frames, creating and applying a filter-dependent flat field (using
the data themselves) and aligning the frames from the positions of
objects within the fields (or if there are insufficient of these to
reliably define a transformation then the telescope offsets written in
the frame headers are used instead), before co-adding them to create a
final image.  After 2000 August 1 we use the {\sc oracdr\_ufti} setup
after applying the {\sc fits2ndf} command from the {\sc starlink} {\sc
convert} package to convert the {\sc fits} files into a merged NDF
format file (between 2001 January 1 and approximately 2001 November 16
we first require to rename the {\sc .fit} file extensions).  The {\sc
oracdr} pipeline is then identical to that run on the earlier data.
For the most part the pipeline reduces the observations with no
intervention using the reduction recipes given in the headers of the
data frames.

The co-added frames from each of the jitter sets in a given filter of a
particular star were then combined together using {\sc starlink} {\sc
ccdpack} routines to give a final frame in each of the $J$-, $H$- and
$K$-bands for each standard star field.  Due to the removal of poor
quality data and problems retrieving data from the archive from
particular nights, the cumulative exposure times for some fields are
slightly below those initially anticipated (Table~1).

In total, data were retrieved for the selected standard stars from over
360 nights from 1999 to 2004, with around 10,000 individual frames
retrieved and processed.

%
%
\begin{figure}
\centerline{\psfig{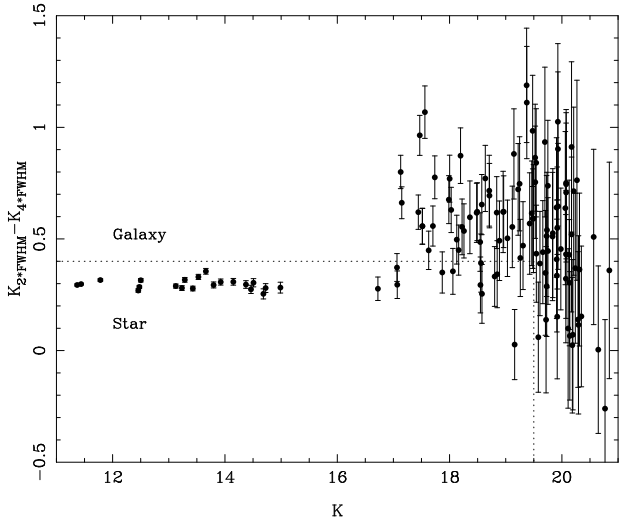}}
\caption{Our criteria used for star-galaxy classification, following
Reid et al.\ (1996). We plot the difference between the $K$-band
magnitudes of all objects measured in apertures with diameters 2$*$ and
4$*$ the FWHM of the FS in that particular field, against their
large-aperture $K$-band magnitudes.  The FS stars, and other bright
stars serendipitously detected in these fields, exhibit a constant
value for $K_{\rm 2*FWHM}-K_{\rm 4*FWHM}$ and we use the mean and
scatter for these to determine a dividing line between
morphologically-classified stars and galaxies -- at $K_{\rm
2*FWHM}-K_{\rm 4*FWHM}=0.40$.  Fainter than $K=19.5$ the measured
magnitudes lack sufficient precision to reliably determine this
classification.  Due to the steeply rising galaxy number counts it is
expected that the vast majority of sources at $K\geq 19.5$ will be
galaxies and hence we class all objects fainter than this as galaxies,
unless they have $K_{\rm 2*FWHM}-K_{\rm 4*FWHM}\leq 0.40$ and colours
consistent with those of the brighter stars in our sample (Fig.~3).
}
\end{figure}

\section{Analysis \& Discussion}

The mean exposure times for the 16 FS in our final sample are 3.2\,ks,
1.7\,ks and 3.7\,ks in $J$-, $H$- and $K$-band respectively. This is
sufficient to reach 5-$\sigma$ point-source limits of $J\sim 21.4$,
$H\sim 20.3$ and $K\sim 20.3$ -- adequate to detect $L^\ast$ galaxies at
$z\sim1$ and beyond.  The median seeing in the $K$-band images is
0.57$''$ as measured from the FS themselves.  The FWHM for each stacked
$K$-band image is reported in Table~1.  The median FWHM in the $J$- and
$H$-bands is 0.70$''$ and 0.67$''$ respectively.  As these are measured
from the co-added images resulting from observations spread over 5 years
these values can be taken to represent firm upper limits on the typical
seeing achieved by UKIRT.

\subsection{Galaxy Catalogues}

Catalogues were constructed from the $K$-band frames, after trimming
these to a $60''\times 60''$ region centred on the FS. To construct
our catalogues we use {\sc SExtractor} (Bertin \& Arnouts 1996) within
{\sc gaia}.  The frames were convolved with a Gaussian kernel with a
FWHM comparable to the seeing, before objects were identified using a
minimum area of 9 pixels each above a threshold of 1-$\sigma$ of the
sky variance.  Catalogues were visually inspected and cleaned of a
small number of faint spurious features.

For our sample we select only those objects with 2$''$-diameter
aperture magnitudes brighter than the 5-$\sigma$ limiting $K$-band
magnitudes listed in Table~1.  These magnitude limits were determined
from the fluctuations in 2$''$-diameter apertures randomly placed on
blank sky regions within the frames.  These estimates suggested that
the typical 5-$\sigma$ depth of our observations is $K\sim 20.3$.  We
quote aperture $K$-band magnitudes for all of our detected sources in
Table~2, the median correction from these to ''total'' magnitudes is
$-0.18\pm 0.04$ as measured from the bright stars in the fields.  We
also note that we have applied no reddening corrections to any of the
magnitudes in this study due to the low reddening in these fields in
the near-infrared.  We measure colours for all objects in
1.0$''$-diameter apertures, sufficient to give representative colours
for the typically compact faint galaxies and stars in our sample.
These are corrected for the seeing differences between the $JHK$ frames
in each individual fields, as measured from the FS star, with typical
aperture corrections of: $\Delta (J-K)=-0.15$ and $\Delta(H-K)=-0.11$.
The colours or 2-$\sigma$ limits listed in Table~2 and used in Figs.~4
\& 5 have had these aperture corrections applied.

The total number of objects detected within 25$''$ radius of the FS
(excluding the FS itself) and brighter than the 5-$\sigma$ limits of
the respective fields, $K=19.7$--20.7, are listed in Table~1.  Out to
25$''$ over the 16 fields, we cover a surveyed area of 8.7 sq.\ arcmin
and detect 111 objects (excluding the FS) brighter than the 5-$\sigma$
limits.

Star-galaxy separation is achieved by comparing the magnitudes within
apertures of 2\,$*$\,FWHM and 4\,$*$\,FWHM (Reid et al.\ 1996).  This
provides a relatively clear differentiation between objects with
star-like and non-star light profiles (Fig.~2).  The FS stars exhibit a
constant offset $K_{\rm 2*FWHM} - K_{\rm 4*FWHM}=0.30\pm0.02$, with
extended galaxies having larger offsets.  Roughly two-thirds of the sample have
$K_{\rm 2*FWHM} - K_{\rm 4*FWHM}>0.40$ which we take as the dividing line in our
classification to $K\leq 19.5$ (Table~2).  Fainter than $K=19.5$ we
have classified all objects as galaxies, unless they have 
$K_{\rm 2*FWHM} - K_{\rm 4*FWHM}\leq 0.40$ and colours consistent with
those of the brighter stars in our sample, $(J-K)\leq 1$
and $(H-K)\leq 0.6$ (Fig.~4), in which case they are classed as stars -- this
effects only 8 objects in our sample.  In total we have 
24 stars (plus the 16 FS) and 87 galaxies.

We list the properties of the 111 objects and the 16 FS stars in our
catalogue in Table~2.  The objects in each field are identified by 
their ranked distance from the FS star and we give their short name,
the position, $K$-band magnitude in our 2$''$-diameter aperture
$(J-K)$ and $(H-K)$ aperture colours or limits, star-galaxy
classification (S or G) and radial distance from the FS star in arcsec.
The astrometry in these images are tied to USNO but due to
the limited fields of view the astrometry is not accurate to
better than 0.5$''$ rms.\footnote{If more precise relative positions are  
required please contact the Authors for the reduced images.}

%
%
\begin{figure}
\centerline{\psfig{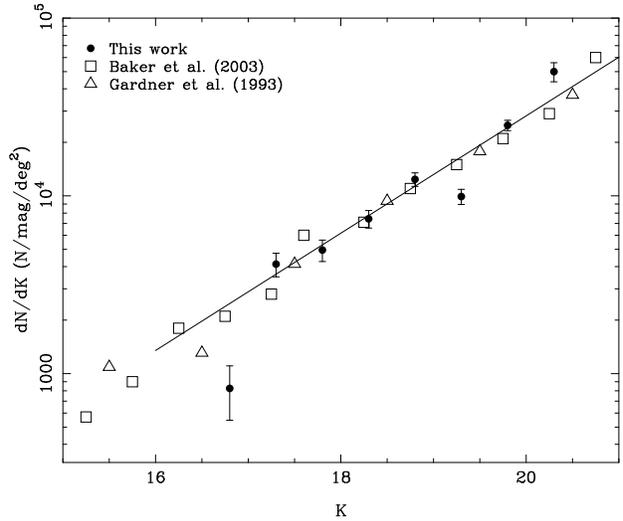}}
\caption{The differential number counts as a function of their
2$''$-diameter aperture magnitudes for sources detected in the
25$''$-radius regions around the 16 FS stars in our sample.  We plot
just the faint-end of our counts, excluding the bright FS stars.  We
correct the number counts for survey area using the 5-$\sigma$
detection limits for each field from Table~1.  The error-bars on our
data points are Poisson and so underestimate the likely variance
resulting from the clustering of galaxies in our small survey fields.
We compare our counts to those from the similar study of Baker et al.\
(2003) and also from Gardner et al.\ (1993), showing generally good
agreement between the various surveys.  We also show the canonical
$\alpha=0.33$ count slope for $K$-band galaxy counts, normalised to fit
our data points.
}
\end{figure}

\subsection{Galaxy Properties}

We show the number counts of sources in our fields (excluding the
bright stars and the FS stars) in Fig.~3.  We compare these to
representative counts from the literature (Gardner et al.\ 1993; Baker
et al.\ 2003) and find good agreement in the overall number of sources
we detect and the slope of the differential number counts, $\alpha \sim
0.3$.  We take this as confirmation that our catalogue is not
significantly incomplete above our typical limiting magnitude, $K\sim
20.3$.

%
%
\begin{figure}
\centerline{\psfig{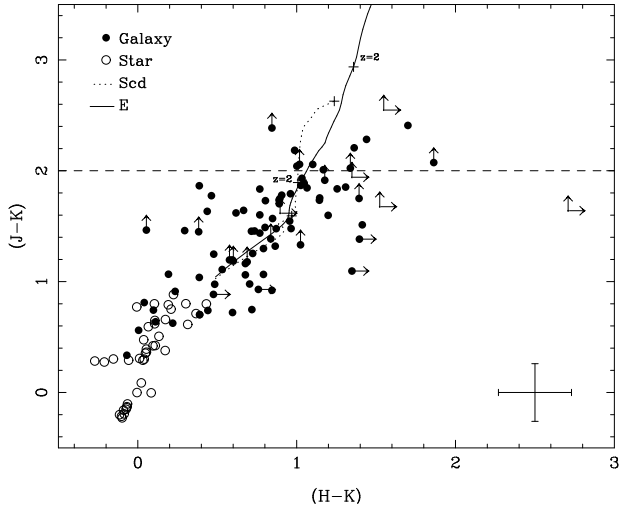}}
\caption{A plot of the distribution of $(J-K)$ and $(H-K)$ colours for
galaxies and stars in our catalogue.  Galaxies and stars are
distinguished by their symbols and we plot 2-$\sigma$ lower limits
where there is no formal detection in the $J$- or $H$-bands.  The stars
show a tight locus and are generally bluer in both $(J-K)$ and $(H-K)$
than the galaxies, which have a broader colour distribution.  We plot
two tracks showing the variation in apparent colour for galaxies with
spectral energy distributions comparable to present-day ellipticals or
Scd galaxies (we have shifted both tracks by $\Delta H=0.2$ to account
for calibration uncertainties between the models and observations).
These start with blue $(J-K)$ and $(H-K)$ colours at $z=0$ and become
increasingly red at higher redshifts, the tracks extend to $z=3$ and we
identify the point corresponding to $z=2$ on both tracks.  The
distribution of galaxy colours broadly spans that expected for passive
or star-forming galaxies at $z\sim 0$--2. The dividing line at
$(J-K)=2$ roughly corresponds to galaxies at $z\gs 2$ as first noted by
Franx et al.\ (2003).  We show the median errors for the colours of our
objects in the lower-right-hand corner.  
}
\end{figure}

In Fig.~4 we show the $(J-K)$--$(H-K)$ colour-colour diagram for all
objects in our fields (including the FS stars).  The figure shows that
those objects classified as stars on the basis of their light profiles
exhibit a tight locus, with relatively blue $(J-K)$ and $(H-K)$
colours, consistent with the measured colours of the FS stars.  The
galaxies have a much broader distribution, extending to redder $(J-K)$
and $(H-K)$ colours.  The reddest galaxies have $(J-K)\gs 2$ or
$(H-K)\gs 1.5$. We plot tracks showing the variation in colour of
galaxies with spectral energy distributions similar to present-day
ellipticals and an Scd galaxy.  These tracks illustrate that the
galaxies with the reddest $(J-K)$ colours are likely to lie at high
redshifts, $z\gs 2$ (Franx et al.\ 2003).  We note that we have 12
galaxies with $(J-K)\geq 2.0$ brighter than $K\sim 20.5$, with 3
galaxies with colours of $(J-K)\geq 2.3$.

Focusing on the 12 $(J-K)\geq 2.0$ galaxies, we find that they are all
well-resolved with FWHM (corrected for seeing) of $\sim 1.4''$ ($\sim
10$\,kpc at $z\sim 2$) and have a median magnitude of $K=19.0$. 
The mean $(H-K)$ colour of these galaxies is $1.3\pm 0.3$, consistent
with the colours expected for $z\sim 2$ galaxies.  These
galaxies are well-situated for AO-assisted studies with a median
separation from the FS star of 13.0$''$.  Half of these galaxies
are found in just two fields, FS\,27 and FS\,34, as expected if
they represent a strongly-clustered population of high-redshift
galaxies (Daddi et al.\ 2004).

\section{Conclusions}

We have undertaken an archival survey for galaxies close to bright
stars which would be suitable for targeting with adaptive optics
imagers and spectrographs on 4--8-m class telescopes.  Our survey
relies on the homogeneous data archive available for the UKIRT
telescope and especially the large quantities of high-quality
calibration imaging taken with the UFTI near-infrared camera.  These
images, of stars from the UKIRT Faint Standards List, provide deep
$JHK$ observations of regions covering the isoplanatic patch around 16
optically-bright, $V\ls 15$ stars.  By co-adding all of the data
available for these 16 stars we survey 25$''$-radius regions around
each star to provide a total area coverage of 8.7 sq.\ arcmins to a
5-$\sigma$ depth of $K\sim 20.3$.

Disregarding the FS guide stars, we detect 111 objects in the 8.7 sq.\
arcmin survey region and exploiting the $\sim 0.5''$ average image
quality from our data, we classify these into stars or galaxies on the
basis of their light profiles.  We find 87 galaxies and 24 stars.  The
$(J-K)$--$(H-K)$ colour-colour distributions for these objects show
that as expected the stars are typically bluer in both $(J-K)$ and
$(H-K)$ than the galaxies.  Moreover, we identify 12 galaxies whose
$(J-K)$ colours exceed $(J-K)=2.0$, consistent with these being
high-redshift galaxies.  These galaxies have $K$-band magnitudes of
$K=18.1$--20.1 and separations from the FS guide star of 4--20$''$ and
hence are very well-suited to AO studies to investigate their
morphologies and spectral properties on sub-kpc scales.

\section*{Acknowledgments}
We thank Malcolm Currie, Brad Cavanagh, Peter Draper Jim Lewis, Nigel
Metcalfe and Mark Swinbank for help. This paper uses data taken with UKIRT
and retrieved from the UKIRT Archive operated by the UK Astronomical
Data Centre.  NMC acknowledges support from a Summer Studentship at
Durham University.  IRS acknowledges support from the Royal Society.

%
%
\begin{table*}
{\small
\begin{center}
\caption{Catalogue of objects in FS fields.
}
\begin{tabular}{lrrccccr}
\noalign{\medskip}
\hline
ID &  R.A.~~~~ & Dec.~~~~ &  $K_{ap}$ & $(J-K)$ & $(H-K)$ & Class & $r$~~ \\ 
    &  \multispan2{ ~(J2000) }  & & & & & ($''$) \\
\noalign{\smallskip}
\hline
FS5-1 & 01~54~34.587 & $-$06~45~59.21 & 12.51$\pm$0.01 & 0.00$\pm$0.01 & $-$0.01$\pm$0.01 & S &  0.0 \cr
FS5-2 & 01~54~34.758 & $-$06~46~04.14 & 19.72$\pm$0.20 &  $\geq$2.06 &  1.02$\pm$0.44 & G &  6.1 \cr
FS5-3 & 01~54~33.961 & $-$06~46~08.53 & 19.56$\pm$0.18 &  1.55$\pm$0.39 &  0.96$\pm$0.44 & G & 13.6 \cr
FS5-4 & 01~54~33.567 & $-$06~45~59.48 & 16.76$\pm$0.04 &  0.00$\pm$0.05 &  0.08$\pm$0.06 & S & 15.8 \cr
FS5-5 & 01~54~35.578 & $-$06~46~02.20 & 19.98$\pm$0.23 &  $\geq$1.62 &  $\geq$0.90 & G & 15.9 \cr
FS5-6 & 01~54~34.284 & $-$06~45~43.21 & 20.03$\pm$0.24 &  1.87$\pm$0.49 &  0.39$\pm$0.35 & G & 17.2 \cr
FS5-7 & 01~54~35.658 & $-$06~45~47.21 & 18.33$\pm$0.09 &  1.60$\pm$0.21 &  0.77$\pm$0.18 & G & 20.9 \cr
FS5-8 & 01~54~33.566 & $-$06~46~12.73 & 17.27$\pm$0.05 &  1.44$\pm$0.13 &  0.77$\pm$0.12 & G & 21.1 \cr
FS5-9 & 01~54~35.356 & $-$06~46~19.52 & 19.77$\pm$0.21 &  1.38$\pm$0.38 &  $\geq$1.39 & G & 24.5 \cr
\noalign{\smallskip}
FS6-1 & 02~30~16.641 & 05~15~50.71 & 13.53$\pm$0.01 & $-$0.13$\pm$0.01 & $-$0.07$\pm$0.01 & S &  0.0 \cr
FS6-2 & 02~30~16.104 & 05~15~54.83 & 19.93$\pm$0.21 &  1.63$\pm$0.52 &  0.44$\pm$0.45 & G &  9.0 \cr
FS6-3 & 02~30~15.972 & 05~16~00.66 & 18.02$\pm$0.07 &  1.64$\pm$0.21 &  0.67$\pm$0.18 & G & 14.1 \cr
FS6-4 & 02~30~17.711 & 05~15~53.69 & 18.61$\pm$0.10 &  1.85$\pm$0.29 &  1.31$\pm$0.35 & G & 16.3 \cr
FS6-5 & 02~30~17.227 & 05~16~11.51 & 20.34$\pm$0.27 &  $\geq$1.46 &  0.05$\pm$0.48 & G & 22.6 \cr
FS6-6 & 02~30~18.151 & 05~15~49.26 & 19.43$\pm$0.15 &  0.89$\pm$0.38 &  $\geq$0.48 & G & 22.6 \cr
FS6-7 & 02~30~16.792 & 05~16~13.55 & 17.46$\pm$0.06 & 0.46$\pm$0.14 &  0.71$\pm$0.12 & G & 22.9 \cr
\noalign{\smallskip}
FS11-1 & 04~52~58.863 & $-$00~14~41.08 & 11.48$\pm$0.00 &  0.09$\pm$0.01 &  0.02$\pm$0.01 & S &  0.0 \cr
FS11-2 & 04~52~58.585 & $-$00~14~42.71 & 18.21$\pm$0.10 & $\geq$2.55 & $\geq$1.55 & G &  4.5 \cr
FS11-3 & 04~52~59.050 & $-$00~14~52.06 & 19.70$\pm$0.24 &  0.80$\pm$0.32 &  0.10$\pm$0.26 & S & 12.0 \cr
FS11-4 & 04~53~00.058 & $-$00~14~37.34 & 14.78$\pm$0.02 &  0.79$\pm$0.03 &  0.19$\pm$0.02 & S & 18.5 \cr
FS11-5 & 04~52~59.721 & $-$00~14~26.47 & 18.70$\pm$0.13 &  2.18$\pm$0.41 &  0.99$\pm$0.25 & G & 20.1 \cr
\noalign{\smallskip}
FS15-1 & 08~51~05.793 & 11~43~47.40 & 12.53$\pm$0.01 &  0.39$\pm$0.01 &  0.05$\pm$0.01 & S &  0.0 \cr
FS15-2 & 08~51~05.886 & 11~43~52.88 & 17.78$\pm$0.07 &  1.20$\pm$0.14 &  0.60$\pm$0.11 & G &  5.7 \cr
FS15-3 & 08~51~04.626 & 11~43~45.98 & 19.73$\pm$0.18 &  1.93$\pm$0.48 &  1.03$\pm$0.35 & G & 17.1 \cr
FS15-4 & 08~51~04.461 & 11~43~51.76 & 18.31$\pm$0.09 &  1.78$\pm$0.22 &  0.91$\pm$0.16 & G & 20.0 \cr
FS15-5 & 08~51~06.284 & 11~43~27.45 & 20.36$\pm$0.26 &  $\geq$1.45 &  0.38$\pm$0.41 & G & 21.2 \cr
FS15-6 & 08~51~04.671 & 11~43~30.89 & 19.98$\pm$0.21 &  1.78$\pm$0.56 &  0.46$\pm$0.36 & G & 23.3 \cr
\noalign{\smallskip}
FS19-1 & 10~33~42.772 & $-$11~41~38.32 & 14.08$\pm$0.01 & $-$0.20$\pm$0.02 & $-$0.12$\pm$0.01 & S &  0.0 \cr
FS19-2 & 10~33~41.807 & $-$11~41~18.79 & 18.66$\pm$0.10 &  0.71$\pm$0.17 &  0.37$\pm$0.15 & S & 24.1 \cr
FS19-3 & 10~33~41.203 & $-$11~41~45.82 & 19.62$\pm$0.17 &  0.70$\pm$0.13 &  0.39$\pm$0.11 & G & 24.2 \cr
FS19-4 & 10~33~41.206 & $-$11~41~47.32 & 17.83$\pm$0.06 &  0.70$\pm$0.13 &  0.39$\pm$0.11 & G & 24.7 \cr
\noalign{\smallskip}
FS20-1 & 11~07~59.958 & $-$05~09~26.43 & 13.70$\pm$0.01 & $-$0.10$\pm$0.01 & $-$0.06$\pm$0.01 & S &  0.0 \cr
FS20-2 & 11~08~00.230 & $-$05~09~24.32 & 20.25$\pm$0.28 &  $\geq$1.33 &  1.02$\pm$0.60 & G &  4.6 \cr
FS20-3 & 11~08~00.535 & $-$05~09~25.07 & 19.34$\pm$0.16 &  2.01$\pm$0.45 &  1.17$\pm$0.35 & G &  8.8 \cr
FS20-4 & 11~08~00.865 & $-$05~09~15.79 & 17.10$\pm$0.05 &  0.88$\pm$0.08 &  0.22$\pm$0.07 & S & 17.3 \cr
FS20-5 & 11~07~59.824 & $-$05~09~47.64 & 20.26$\pm$0.29 &  1.18$\pm$10.01 &  0.69$\pm$0.63 & G & 21.4 \cr
\noalign{\smallskip}
FS21-1 & 11~37~05.102 & 29~47~58.37 & 13.31$\pm$0.01 & $-$0.16$\pm$0.01 & $-$0.09$\pm$0.01 & S &  0.0 \cr
FS21-2 & 11~37~04.632 & 29~48~00.19 & 20.40$\pm$0.27 &  $\geq$1.39 &  0.84$\pm$0.54 & G &  6.3 \cr
FS21-3 & 11~37~05.559 & 29~47~48.45 & 20.09$\pm$0.23 &  $\geq$2.02 &  1.34$\pm$0.47 & G & 11.6 \cr
FS21-4 & 11~37~04.282 & 29~48~05.80 & 18.79$\pm$0.11 &  1.89$\pm$0.32 &  1.05$\pm$0.23 & G & 13.0 \cr
FS21-5 & 11~37~05.051 & 29~48~11.87 & 19.68$\pm$0.18 &  $\geq$2.39 &  0.84$\pm$0.31 & G & 13.5 \cr
FS21-6 & 11~37~04.316 & 29~48~08.38 & 17.59$\pm$0.06 &  1.25$\pm$0.16 &  0.72$\pm$0.13 & G & 14.3 \cr
FS21-7 & 11~37~05.329 & 29~48~15.93 & 17.85$\pm$0.07 &  1.48$\pm$0.18 &  0.87$\pm$0.14 & G & 17.8 \cr
FS21-8 & 11~37~03.698 & 29~47~54.10 & 19.11$\pm$0.13 &  1.48$\pm$0.28 &  0.96$\pm$0.24 & G & 18.7 \cr
FS21-9 & 11~37~03.640 & 29~48~00.78 & 19.59$\pm$0.17 &  1.73$\pm$0.43 &  0.80$\pm$0.31 & G & 19.1 \cr
FS21-10 & 11~37~03.389 & 29~48~00.94 & 18.57$\pm$0.10 &  1.70$\pm$0.25 &  0.89$\pm$0.19 & G & 22.4 \cr
FS21-11 & 11~37~03.375 & 29~47~51.54 & 19.37$\pm$0.15 & 1.85$\pm$0.42 &  1.07$\pm$0.33 & G & 23.4 \cr
\noalign{\smallskip}
FS23-1 & 13~41~43.626 & 28~29~51.47 & 12.54$\pm$0.01 &  0.59$\pm$0.01 &  0.07$\pm$0.01 & S &  0.0 \cr
FS23-2 & 13~41~43.959 & 28~29~43.48 & 19.16$\pm$0.14 &  1.84$\pm$0.40 &  0.77$\pm$0.28 & G &  9.1 \cr
FS23-3 & 13~41~42.926 & 28~29~55.22 & 20.18$\pm$0.25 &  0.27$\pm$0.32 & $-$0.21$\pm$0.29 & S &  9.7 \cr
FS23-4 & 13~41~44.110 & 28~30~00.28 & 20.07$\pm$0.23 &  0.61$\pm$0.34 &  0.31$\pm$0.33 & S & 10.8 \cr
FS23-5 & 13~41~42.698 & 28~29~46.71 & 20.12$\pm$0.24 &  0.51$\pm$0.31 &  0.13$\pm$0.28 & S & 13.2 \cr
FS23-6 & 13~41~43.922 & 28~30~04.27 & 20.37$\pm$0.28 &  0.75$\pm$0.52 &  0.72$\pm$0.59 & G & 13.2 \cr
FS23-7 & 13~41~42.557 & 28~29~51.71 & 19.89$\pm$0.21 &  0.56$\pm$0.32 &  0.00$\pm$0.28 & G & 13.9 \cr
FS23-8 & 13~41~43.345 & 28~30~05.20 & 20.44$\pm$0.29 &  $\geq$1.20 &  0.58$\pm$0.53 & G & 14.0 \cr
FS23-9  & 13~41~43.501 & 28~30~07.12 & 17.92$\pm$0.07 &  0.31$\pm$0.11 &  0.01$\pm$0.10 & S & 15.6 \cr
FS23-10 & 13~41~44.967 & 28~29~46.80 & 20.40$\pm$0.28 &  0.36$\pm$0.38 &  0.05$\pm$0.36 & S & 18.1 \cr
FS23-11 & 13~41~42.230 & 28~29~52.22 & 19.82$\pm$0.20 &  0.62$\pm$0.35 &  0.22$\pm$0.33 & G & 18.2 \cr
FS23-12 & 13~41~44.196 & 28~30~08.94 & 18.21$\pm$0.08 &  1.75$\pm$0.22 &  0.89$\pm$0.16 & G & 19.0 \cr
FS23-13 & 13~41~45.121 & 28~29~49.74 & 20.04$\pm$0.22 &  0.28$\pm$0.30 & $-$0.27$\pm$0.27 & S & 19.7 \cr
FS23-14 & 13~41~43.937 & 28~29~31.04 & 14.55$\pm$0.01 &  0.47$\pm$0.02 &  0.04$\pm$0.02 & S & 20.7 \cr
FS23-15 & 13~41~42.253 & 28~30~02.52 & 20.07$\pm$0.23 &  1.46$\pm$0.56 &  0.30$\pm$0.40 & G & 20.8 \cr
FS23-16 & 13~41~42.440 & 28~30~05.79 & 18.08$\pm$0.08 &  0.30$\pm$0.12 & $-$0.15$\pm$0.11 & S & 20.9 \cr
FS23-17 & 13~41~42.173 & 28~30~00.78 & 20.42$\pm$0.28 &  0.91$\pm$0.48 &  0.23$\pm$0.42 & G & 21.0 \cr
FS23-18 & 13~41~42.010 & 28~29~46.49 & 17.11$\pm$0.05 &  0.29$\pm$0.07 & $-$0.06$\pm$0.07 & S & 21.7 \cr
FS23-19 & 13~41~44.354 & 28~30~11.15 & 19.99$\pm$0.22 &  0.81$\pm$0.48 &  0.04$\pm$0.41 & G & 21.9 \cr
\hline
\end{tabular}
\end{center}}
\end{table*}
\newpage

\setcounter{table}{1}
\begin{table*}
{\small
\begin{center}
\caption{cont.
}
\begin{tabular}{lrrccccr}
\noalign{\medskip}
\hline
ID &  R.A.~~~~ & Dec.~~~~ &  $K_{ap}$ & $(J-K)$ & $(H-K)$ & Class & $r$~~ \\ 
    &  \multispan2{ ~(J2000) }  & & & & & ($''$) \\
\noalign{\smallskip}
\hline

\noalign{\smallskip}
FS27-1 & 16~40~41.615 & 36~21~13.23 & 13.31$\pm$0.01 &  0.35$\pm$0.01 &  0.05$\pm$0.01 & S &  0.0 \cr
FS27-2 & 16~40~41.306 & 36~21~15.50 & 19.94$\pm$0.19 & 1.16$\pm$0.40 & 0.66$\pm$0.40 & G &  4.2 \cr
FS27-3 & 16~40~41.499 & 36~21~23.23 & 19.67$\pm$0.17 &  1.46$\pm$0.38 &  0.73$\pm$0.29 & G & 10.0 \cr
FS27-4 & 16~40~40.771 & 36~21~06.97 & 19.07$\pm$0.12 &  0.80$\pm$0.19 &  0.43$\pm$0.17 & S & 11.9 \cr
FS27-5 & 16~40~41.224 & 36~21~01.40 & 19.93$\pm$0.19 &  1.32$\pm$0.38 &  0.86$\pm$0.32 & G & 12.7 \cr
FS27-6 & 16~40~40.475 & 36~21~14.11 & 20.13$\pm$0.22 &  $\geq$2.07 &  1.86$\pm$0.52 & G & 13.7 \cr
FS27-7 & 16~40~40.524 & 36~21~18.33 & 18.65$\pm$0.10 &  0.38$\pm$0.15 &  0.17$\pm$0.14 & S & 14.0 \cr
FS27-8 & 16~40~40.633 & 36~21~02.73 & 20.14$\pm$0.22 &  $\geq$1.68 &  $\geq$1.52 & G & 15.8 \cr
FS27-9 & 16~40~41.650 & 36~20~57.14 & 20.09$\pm$0.21 &  2.04$\pm$0.63 &  1.00$\pm$0.42 & G & 16.1 \cr
FS27-10 & 16~40~40.922 & 36~21~27.54 & 17.90$\pm$0.07 &  1.87$\pm$0.19 &  1.03$\pm$0.13 & G & 16.4 \cr
FS27-11 & 16~40~41.148 & 36~20~57.67 & 20.38$\pm$0.25 &  0.74$\pm$0.45 &  0.10$\pm$0.37 & G & 16.5 \cr
FS27-12 & 16~40~42.199 & 36~21~28.34 & 18.59$\pm$0.10 &  2.28$\pm$0.35 &  1.44$\pm$0.35 & G & 16.6 \cr
FS27-13 & 16~40~42.975 & 36~21~09.96 & 19.38$\pm$0.15 &  0.98$\pm$0.35 &  0.70$\pm$0.32 & G & 16.8 \cr
FS27-14 & 16~40~42.268 & 36~20~57.96 & 20.49$\pm$0.26 &  1.04$\pm$0.42 &  0.39$\pm$0.34 & G & 17.2 \cr
FS27-15 & 16~40~41.519 & 36~21~30.55 & 19.71$\pm$0.17 &  0.64$\pm$0.29 &  0.11$\pm$0.24 & G & 17.2 \cr
FS27-16 & 16~40~42.377 & 36~21~30.12 & 14.53$\pm$0.01 &  0.75$\pm$0.02 &  0.21$\pm$0.02 & S & 19.2 \cr
FS27-17 & 16~40~43.277 & 36~21~11.65 & 19.98$\pm$0.20 &  1.30$\pm$0.43 &  0.79$\pm$0.36 & G & 20.1 \cr
FS27-18 & 16~40~40.510 & 36~20~57.15 & 20.38$\pm$0.25 &  0.66$\pm$0.40 &  0.17$\pm$0.35 & S & 20.9 \cr
FS27-19 & 16~40~40.057 & 36~21~03.21 & 18.69$\pm$0.10 &  0.42$\pm$0.16 &  0.09$\pm$0.14 & S & 21.2 \cr
FS27-20 & 16~40~42.954 & 36~20~59.29 & 18.95$\pm$0.12 &  0.42$\pm$0.18 &  0.11$\pm$0.16 & S & 21.4 \cr
FS27-21 & 16~40~43.490 & 36~21~03.28 & 17.40$\pm$0.06 &  1.74$\pm$0.15 &  0.89$\pm$0.11 & G & 24.8 \cr
\noalign{\smallskip}
FS29-1 & 21~52~25.358 & 02~23~19.34 & 13.45$\pm$0.01 & $-$0.14$\pm$0.01 & $-$0.07$\pm$0.01 & S &  0.0 \cr
FS29-2 & 21~52~24.740 & 02~23~15.60 & 18.81$\pm$0.11 &  1.49$\pm$0.27 &  0.80$\pm$0.20 & G &  9.9 \cr
FS29-3 & 21~52~24.933 & 02~23~05.94 & 20.40$\pm$0.27 &  0.74$\pm$0.42 &  0.44$\pm$0.37 & G & 14.8 \cr
FS29-4 & 21~52~24.301 & 02~23~26.39 & 18.10$\pm$0.07 &  1.79$\pm$0.21 &  0.96$\pm$0.14 & G & 17.2 \cr
FS29-5 & 21~52~25.420 & 02~23~39.32 & 14.75$\pm$0.02 &  0.65$\pm$0.03 &  0.11$\pm$0.02 & S & 20.0 \cr
FS29-6 & 21~52~26.697 & 02~23~20.29 & 19.81$\pm$0.18 &  1.60$\pm$0.51 &  1.20$\pm$0.42 & G & 20.1 \cr
FS29-7 & 21~52~25.200 & 02~22~57.37 & 19.74$\pm$0.18 &  0.34$\pm$0.33 & $-$0.07$\pm$0.29 & G & 22.1 \cr
FS29-8 & 21~52~23.848 & 02~23~09.88 & 19.59$\pm$0.16 &  1.11$\pm$0.44 &  0.53$\pm$0.35 & G & 24.4 \cr
FS29-9 & 21~52~24.088 & 02~23~03.60 & 20.08$\pm$0.22 &  $\geq$1.18 &  0.60$\pm$0.46 & G & 24.6 \cr
\noalign{\smallskip}
FS31-1 & 23~12~21.653 & 10~47~05.06 & 14.17$\pm$0.01 & $-$0.23$\pm$0.01 & $-$0.10$\pm$0.01 & S &  0.0 \cr
FS31-2 & 23~12~20.987 & 10~47~05.85 & 19.29$\pm$0.14 &  1.07$\pm$0.25 &  0.19$\pm$0.18 & G &  9.8 \cr
FS31-3 & 23~12~22.614 & 10~47~05.77 & 18.82$\pm$0.11 &  0.62$\pm$0.16 &  0.11$\pm$0.14 & S & 14.3 \cr
FS31-4 & 23~12~22.758 & 10~47~12.94 & 19.06$\pm$0.12 &  1.25$\pm$0.26 &  0.48$\pm$0.20 & G & 18.1 \cr
FS31-5 & 23~12~22.593 & 10~46~52.65 & 19.61$\pm$0.16 &  1.62$\pm$0.48 &  0.62$\pm$0.35 & G & 18.6 \cr
FS31-6 & 23~12~21.591 & 10~47~25.34 & 20.12$\pm$0.22 &  $\geq$1.75 &  1.39$\pm$0.52 & G & 20.3 \cr
\noalign{\smallskip}
FS32-1 & 23~16~12.400 & $-$01~50~34.70 & 13.81$\pm$0.01 &  $-$0.19$\pm$0.01 & $-$0.09$\pm$0.01 & S &  0.0 \cr
FS32-2 & 23~16~11.741 & $-$01~50~22.81 & 19.90$\pm$0.23 &  $\geq$1.94 &  $\geq$1.35 & G & 15.8 \cr
FS32-3 & 23~16~12.661 & $-$01~50~16.37 & 15.00$\pm$0.02 &  0.80$\pm$0.03 &  0.30$\pm$0.02 & S & 19.1 \cr
FS32-4 & 23~16~11.130 & $-$01~50~28.41 & 20.03$\pm$0.25 &  0.77$\pm$0.29 & $-$0.01$\pm$0.25 & S & 20.5 \cr
\noalign{\smallskip}
FS33-1 & 12~57~02.337 & 22~01~52.69 & 14.44$\pm$0.01 & $-$0.21$\pm$0.02 & $-$0.10$\pm$0.02 & S &  0.0 \cr
FS33-2 & 12~57~01.833 & 22~01~54.55 & 20.47$\pm$0.29 &  0.93$\pm$0.61 &  $\geq$0.76 & G &  7.2 \cr
FS33-3 & 12~57~02.842 & 22~01~45.43 & 18.64$\pm$0.10 &  1.84$\pm$0.29 &  1.25$\pm$0.22 & G & 10.1 \cr
FS33-4 & 12~57~02.173 & 22~02~10.40 & 19.53$\pm$0.16 &  0.92$\pm$0.27 &  0.84$\pm$0.26 & G & 17.8 \cr
FS33-5 & 12~57~03.893 & 22~01~52.62 & 20.37$\pm$0.27 &  1.10$\pm$0.48 &  $\geq$1.35 & G & 21.7 \cr
FS33-6 & 12~57~02.863 & 22~01~31.83 & 20.17$\pm$0.24 &  $\geq$1.92 &  1.18$\pm$0.44 & G & 22.1 \cr
\noalign{\smallskip}
FS34-1 & 20~42~34.713 & $-$20~04~35.62 & 13.15$\pm$0.01 & $-$0.15$\pm$0.01 & $-$0.07$\pm$0.01 & S &  0.0 \cr
FS34-2 & 20~42~34.019 & $-$20~04~32.30 & 18.54$\pm$0.10 &  2.41$\pm$0.37 &  1.70$\pm$0.26 & G & 10.2 \cr
FS34-3 & 20~42~35.042 & $-$20~04~25.66 & 18.91$\pm$0.13 &  1.76$\pm$0.34 &  1.14$\pm$0.25 & G & 11.0 \cr
FS34-4 & 20~42~35.427 & $-$20~04~42.50 & 19.54$\pm$0.19 &  $\geq$1.64 &  $\geq$2.71 & G & 12.3 \cr
FS34-5 & 20~42~35.832 & $-$20~04~37.38 & 18.73$\pm$0.11 &  2.21$\pm$0.42 &  1.36$\pm$0.29 & G & 15.9 \cr
FS34-6 & 20~42~34.603 & $-$20~04~54.71 & 19.90$\pm$0.24 &  0.72$\pm$0.45 &  0.60$\pm$0.47 & G & 19.1 \cr
FS34-7 & 20~42~33.367 & $-$20~04~31.43 & 19.15$\pm$0.14 &  0.98$\pm$0.26 &  0.48$\pm$0.22 & G & 19.3 \cr
FS34-8 & 20~42~35.564 & $-$20~04~19.87 & 18.07$\pm$0.08 &  2.06$\pm$0.26 &  1.10$\pm$0.16 & G & 19.8 \cr
FS34-9 & 20~42~34.994 & $-$20~04~57.02 & 19.85$\pm$0.23 &  1.06$\pm$0.40 &  0.79$\pm$0.38 & G & 21.8 \cr
FS34-10 & 20~42~33.470 & $-$20~04~51.34 & 19.01$\pm$0.13 &  1.16$\pm$0.26 &  0.68$\pm$0.21 & G & 23.4 \cr
\noalign{\smallskip}
FS127-1 & 10~06~28.916 & 41~01~24.97 & 11.85$\pm$0.00 &  0.30$\pm$0.01 &  0.04$\pm$0.01 & S &  0.0 \cr
FS127-2 & 10~06~27.378 & 41~01~17.66 & 20.37$\pm$0.31 &  1.57$\pm$0.53 &  0.85$\pm$0.56 & G & 18.8 \cr
\noalign{\smallskip}
FS131-1 & 12~14~25.403 & 35~35~54.42 & 11.51$\pm$0.00 &  0.29$\pm$0.01 &  0.03$\pm$0.01 & S &  0.0 \cr
FS131-2 & 12~14~24.421 & 35~36~01.85 & 19.41$\pm$0.18 &  1.51$\pm$0.33 &  1.41$\pm$0.38 & G & 14.0 \cr
FS131-3 & 12~14~26.001 & 35~36~14.32 & 18.37$\pm$0.09 &  1.73$\pm$0.25 &  1.14$\pm$0.22 & G & 21.2 \cr
\hline
\end{tabular}

\end{center}}
\end{table*}

\end{document}